\documentclass[aps,prd,preprintnumbers,showpacs]{revtex4}
\usepackage[dvips]{graphicx}
\begin{document}

%
%

\eprint{Nisho-1-2014}
\title{Monopole Production and Glasma Decay}
\author{Aiichi Iwazaki}
\affiliation{International Economics and Politics, Nishogakusha University,\\ 
6-16 3-bantyo Chiyoda Tokyo 102-8336, Japan.}   
\date{Jun. 7, 2014}
\begin{abstract}
It has been discussed that quark gluon plasma is produced following the rapid decay of coherent color gauge fields
generated immediately after high energy heavy ion collisions. But there are no convincing mechanisms known
for the rapid decay of the gauge fields which satisfy phenomenological
constraints on their lifetime.
We show by using classical statistical field theory that the gauge fields rapidly decay into
magnetic monopoles.
For comparison, we show how fast they decay into 
Nielsen-Olesen unstable modes. We find that
the rapid decay of the glasma is caused by the monopole production.  
\end{abstract}
\hspace*{0.3cm}
\pacs{12.38.-t, 12.38.Mh, 25.75.-q, 14.80.Hv \\
Quark Gluon Plasma, Monopoles, Color Glass Condensate}
\hspace*{1cm}

\maketitle

\section{introduction}
\label{1}
One of the most significant problems
in high energy heavy ion collisions
is how coherent color gauge fields\cite{cgc,cgc1} ( here we call them glasma ) 
generated immediately after the collisions decay into quark gluon plasma (QGP). 
In particular, the fields rapidly decay into thermalized QGP
within a time less than $1$fm/c\cite{hirano}.
The problem is what mechanism causes such a rapid decay of the glasma.

The presence of such coherent gauge fields has been
shown using a model of color glass condensate\cite{cgc,cgc1}.
They are color electric and magnetic fields pointed
in the longitudinal direction of the collisions. 
They are uniform in the longitudinal direction, while they
vary in the transverse directions. Thus, we may think that they exist in the form of 
flux tubes with various widths. 
The typical width is
given by $Q_s^{-1}$; $Q_s$ is saturation momentum in the collisions.
Furthermore, the typical values of the gauge field strength are 
given by $Q_s^2$.

Classical instabilities of such gauge fields have been shown 
in numerical simulations\cite{berges,berges1,kunihiro1,kunihiro2,ven,fuku}. 
It has been found that small 
fluctuations added to the gauge fields grow exponentially.
Consequently, the longitudinal pressure of gauge fields or the distance between
two adjacent classical trajectories in the space of gauge fields grow exponentially. 
The presence of the instabilities implies
 that the gauge fields decay with entropy production\cite{kunihiro1}
and indicates that QGP is eventually produced. However,
the lifetime of the gauge field given by the instabilities
has been shown to be much longer than
$1$fm/c. 


The instabilities are 
Nielsen-Olesen instability\cite{nielsen,instability,instability1,instability2}. 
The authors\cite{berges2}
have demonstrated numerically in detail that
they are Nielsen-Olesen instabilities, not Weibel instabilities.
Nielsen and Olesen have shown in their original paper\cite{nielsen}
that there exist unstable modes growing exponentially with time $\sim \exp(\gamma t)$ 
under homogeneous color magnetic field $B$.   
Their growth rate $\gamma$ is given 
by $\sqrt{gB}$; $g \,\,( >0 )$ is a gauge coupling constant.
On the other hand, the color magnetic fields in the glasma are inhomogeneous. Under such color magnetic fields
unstable modes grow more slowly than those under the homogeneous color magnetic field.
Indeed, $\gamma$ is given by 
square root $\sqrt{gB_{\rm{eff}}} $ of an effective magnetic field $gB_{\rm{eff}}$, 
which is
much smaller than $Q_s^2$, e.g. $\sqrt{gB_{\rm{eff}}}\sim 0.2Q_s$. 
Furthermore, the growth rates\cite{ven,fuku, berges1} in expanding glasma are much smaller than
those in non-expanding glasma. 
Therefore, 
the glasma develops the instabilities with its lifetime 
longer than 
$1$fm/c. 
This contradicts
the phenomenological analysis,
which shows that thermalized QGP is realized within a time less than $1$fm/c.
We must find a new mechanism for the rapid decay of the glasma.

 
Magnetic monopoles\cite{coleman,t,t1} have been discussed to be essential ingredients\cite{ie,s} for
quark confinement. According to the picture of quark confinement,
monopole condensation realizes a confining vacuum of dual superconductor\cite{t,t1,dual}:
In the dual superconductor color electric field is squeezed into a flux tube 
between a quark and an anti-quark.
That is, the monopoles are spontaneously produced in a perturbative vacuum and
condense so that the real confining vacuum is realized.

Their presence as well as role for the quark confinement
have been extensively studied in lattice gauge theories\cite{dual,koma,max,max1,max2}.
Effective models of the dual superconductors
have also been explored\cite{dual,koma} by mainly using the lattice gauge theories as well as
continuous gauge theories\cite{kondo}.
The models are defined using the complex scalar field of the monopoles 
with their ``imaginary mass" just like
the Higgs field. 
Up to now, they have been only analyzed from theoretical point of view 
in order to see the properties of
the confining vacuum.
Because the monopoles have color magnetic charges $\sim 1/g$, 
they can be produced under the color magnetic fields of the glasma. 
Thus, we are tempted to ask how their production affects the decay of the glasma.
As long as we know, it is first time to apply the models
to the phenomenological analysis, in particular, in high energy heavy ion collisions.
Although it is not obvious that the models are applicable to the analysis,
the use of the model in the analysis would be interesting attempt. 



In this paper assuming an effective model of the monopoles in SU(2) gauge theory
we show that the glasma rapidly decays into the magnetic monopoles.
The monopole production arises owing to the Schwinger mechanism\cite{sch}
under background color magnetic fields of the glasma. We calculate
the production rate of the monopoles and 
their back reaction to the color magnetic fields. 
The production rate is obtained by using recent result\cite{ita} concerning the production rate of  
Nielsen-Olesen unstable modes. But, the result does not include
the back reaction of the monopoles.
On the other hand, by using
classical statistical field theory\cite{csft, csft1, csft2} 
we can include the back reaction.
We find that the lifetime of the color magnetic fields can be much shorter than $1$fm/c. 
For comparison, we also calculate the back reaction of 
the Nielsen-Olesen unstable modes using the classical statistical field theory.
We find that the glasma mainly decays into the magnetic monopoles. 
Although we only treat non-expanding glasma,
our result does hold even in expanding glasma.

In next section(\ref{2}), we give a brief review of 
Nielsen-Olesen unstable modes and present an effective model of the modes. 
In section(\ref{3}), we present an effective model of monopoles and show 
that the glasma
predominantly decays into the monopoles, not into
the Nielsen-Olesen unstable modes. Discussions and conclusions follow
in section(\ref{4})

\section{Nielsen-Olesen Instability in Inhomogeneous Magnetic Fields}
\label{2}

First, we briefly review the Nielsen-Olesen unstable modes
found in the previous numerical
simulations. 
In particular, 
we present an effective model of the Nielsen-Olesen unstable modes
arising under ``inhomogeneous" color magnetic field.
The model is used to show how fast the color electric fields
decay into the unstable modes. 
Hereafter, we denote color magnetic ( electric ) field simply
by magnetic ( or electric ) field.

We consider SU(2) gauge theory with
the background electric and magnetic fields given by
$\vec{E}_a=\delta_{a,3}\vec{E}$ and $\vec{B}_a=\delta_{a,3}\vec{B}$.
( Field strength $gB$ or $gE$ of the fields are of the order of $Q_s^2$ in the glasma. )
Here we assume for simplicity that these fields point in the direction of 
the third axis in color space.
The fields are described by the ``electromagnetic" gauge fields $A_{\mu}\equiv A_{\mu}^{a=3}$.
Under the background gauge fields, the fields $\Phi_{\mu}\equiv (A_{\mu}^1+iA_{\mu}^2)/\sqrt{2}$ 
perpendicular to $A_{\mu}^3$ behave
as charged vector fields. When we represent SU(2) gauge fields $A_{\mu}^a$ using 
the variables $A_{\mu}$ and $\Phi_{\mu}$, Lagrangian of SU(2) gauge fields can be
written in the following,

\begin{equation}
L=-\frac{1}{4}F_{\mu,\nu}^2-\frac{1}{2}|D_{\mu}\Phi_{\nu}-D_{\nu}\Phi_{\mu}|^2
-ig(\partial_{\mu}A_{\nu}-\partial_{\nu}A_{\mu})\Phi^{\mu}\Phi^{\dagger \nu}
+\frac{g^2}{4}(\Phi_{\mu}^{\dagger}\Phi_{\nu}-\Phi_{\nu}^{\dagger}\Phi_{\mu})^2,
\end{equation} 
with $F_{\mu,\nu}=\partial_{\mu}A_{\nu}-\partial_{\nu}A_{\mu}$ and 
$D_{\mu}=\partial_{\mu}+igA_{\mu}$.
We understand that the fields $\Phi_{\mu}$ represent charged vector fields
with the anomalous magnetic moment represented by the term 
$-ig(\partial_{\mu}A_{\nu}-\partial_{\nu}A_{\mu})\Phi^{\mu}\Phi^{\dagger \nu}$. 
This term gives rise to the instability of the background magnetic field.

In order to see it, assuming the magnetic field $\vec{B}=(0,0,B)$,
we write down the Hamiltonian of the field $\Phi_i$,

\begin{eqnarray}
H &=&\int d^3x \Bigr( |\partial_t\Phi_i|^2+\frac{1}{2}|D_i\Phi_j-D_j\Phi_i|^2
+igF_{i,j}\Phi_i\Phi_j^{\dagger}\Bigr) \nonumber \\
 &=&\int d^3x \Bigl(|\partial_t\Phi_i|^2+|D_i\Phi_j|^2+2igF_{i,j}\Phi_i\Phi_j^{\dagger}\Bigr) \nonumber \\
 &=&\int d^3x \Bigl(|\partial_t\Phi_+|^2+|\partial_t\Phi_-|^2+
|D_i\Phi_+|^2+|D_i\Phi_-|^2-2gB|\Phi_+|^2+2gB|\Phi_-|^2 \Bigr),
\end{eqnarray}
with $B=F_{1,2}$ and $\Phi_{\pm}\equiv(\Phi_1 \pm i\Phi_2)/\sqrt{2}$,
where we neglected the interaction terms $\sim g^2\Phi_i^4$ and 
the surface terms in partial integrations. 
We used a gauge condition $\Phi_0=0$ as well as a condition $D_i\Phi_i=0$. 
We also neglected the field $\Phi_3$ irrelevant to our discussion, which is determined in terms of
$\Phi_{\pm}$ via the
condition $D_i\Phi_i=0$.
The third term in the first equation represents the anomalous magnetic moment
of the field $\Phi_i$. 

For example, in the presence of a homogeneous background magnetic field 
the particles represented by the fields $\Phi_{\pm}$
occupy the Landau levels denoted by integer $N\ge 0$.
Their energies are 
given by $E_N=\sqrt{2gB(N+1/2)\mp 2gB+p_3^2}$, where $\pm$ denotes magnetic moment
parallel ( $-$ ) or anti-parallel ( $+$ ) to $\vec{B}$,
and $p_3$ denotes a momentum component parallel to $\vec{B}$.
( The term $\mp 2gB$ in $E_N$ comes from 
the term $2igF_{i,j}\Phi_i\Phi_j^{\dagger}=\mp 2gB|\Phi_{\pm}|^2$ ).
Among them the energies of the states 
in the lowest Landau level ( $N=0$ ) 
with the magnetic moment parallel to $\vec{B}$ 
can be imaginary; $E_{N=0}=\sqrt{p_3^2-gB}$ when $p_3^2<gB$. 
Thus, the modes 
with the imaginary energies exponentially increase or decrease with time;
$\Phi_i\propto \exp(-iE_{N=0}t)=\exp(\pm |E_{N=0}|t)=\exp(\pm |\sqrt{gB-p_3^2}|\,t)$.
The modes are called as Nielsen-Olesen
unstable modes. In particular, the mode with $p_3=0$ increases most rapidly
with the growth rate $\sqrt{gB}$. We note that the modes with the energies $E_{N=0}=\sqrt{p_3^2-gB}$
become stable when their momentum is sufficiently large such as $p_3^2>gB$.

The ``kinetic energy" $|D_i\Phi_+|^2$ of the states in the 
lowest Landau level is 
given by $+gB$ as usual, while the ``potential energy" $-2gB|\Phi_+|^2$ given by the anomalous magnetic moment
is negative, that is given by $-2gB$.
Thus, the ``total energy" $E^2$ is given such that $gB-2gB=-gB<0$. This leads to  
the imaginary energy $E=\sqrt{-gB}$. Obviously, the anomalous magnetic moment
causes the instability.


The homogeneous magnetic field is not realistic. 
Magnetic fields in the glasma exist in the form of flux tubes.
Then, the ``potential" ( $-2gB$ ) may
vary in space, in other words, 
the potential can be negative or positive.  
There are still unstable modes even in such a potential, 
although their growth rate is much smaller than 
the typical energy scale $\sqrt{|gB|}$ of the potential. 
Such unstable modes are represented by either $\Phi_+$ or $\Phi_-$ 
for which the average ``potential" is negative.
This is because for example, even if the ``potential" $-2gB$ ( $>0$ ) for $\Phi_+$  is positive in a spatial region, 
there exist gauge fields $\Phi_-$
for which the ``potential" $+2gB$ is negative in the region. 
Thus, even if the spatial average of the ``potential" $-2gB$ for $\Phi_+$
is positive, the average of the ``potential" $2gB$ for $\Phi_-$ is negative.
In this example, $\Phi_-$ represents unstable modes.
Therefore, there are unstable modes even under inhomogeneous $B$, which are represented by 
either $\Phi_+$ or $\Phi_-$.

The typical energy scale of the ``potential" is given by the square of the saturation momentum $Q_s^2$,
but the average depth of the ``potential" is not of the order of $Q_s^2$, but much less than $Q_s^2$.
Hence the growth rates $\gamma$ of the unstable modes 
growing as $\Phi\sim \exp(\gamma t) $ are much smaller than $Q_s$. 
We should remember that
the numerical simulations\cite{berges,berges1,kunihiro1,ven,fuku}
have shown much smaller growth rate $0.1Q_s\sim 0.2Q_s $ than
the typical energy scale $Q_s$.

The small growth rate can be
represented 
by using effective homogeneous magnetic field $gB_{\rm{eff}}$ ($\ll Q_s^2$);
 $\gamma=\sqrt{gB_{\rm{eff}}}$.
Under the effective magnetic field,
the Nielsen-Olesen unstable modes exponentially grow, i.e. $\exp(t\sqrt{gB_{\rm{eff}}})$.
Similarly, the average effect of inhomogeneous electric field 
in the glasma can be described 
by using effective homogeneous electric field $E_{\rm{eff}}$.
Therefore, in order to discuss back reactions of the unstable modes to the electric field,
we may consider the following effective Lagrangian of the 
Nielsen-Olesen unstable modes $\phi_{NO}\equiv (\Phi_1+i\Phi_2)/\sqrt{2}$,

\begin{eqnarray}
\label{NO}
L_{\rm{eff}} &=&|(\partial_{\mu}+igA_{\rm{eff},\mu})\phi_{NO}|^2+2gB_{\rm{eff}}|\phi_{NO}|^2 \nonumber \\
 &=&|\partial_t\phi_{NO}|^2-|(\partial_3+igA_{\rm{eff},3})\phi_{NO}|^2
+gB_{\rm{eff}}|\phi_{NO}|^2,
\end{eqnarray} 
where we have taken into account only the states in the lowest Landau level and 
neglected the interaction terms $\sim g^2\phi_{NO}^4$.
The effective homogeneous background magnetic field ( electric field ) is given by
$B_{\rm{eff}}=\partial_1A_{\rm{eff},2}-\partial_2A_{\rm{eff},1}$ ( $E_{\rm{eff}}=\partial_0 A_{\rm{eff},3}$).
Both $B_{\rm{eff}}$ and $E_{\rm{eff}}$ are of the same order of magnitude 
in the initial stage when they are produced. But, subsequently the electric field decreases
with the production of the modes $\phi_{NO}$, while the magnetic field does not.

In the next section, using the Lagrangian we show how fast the electric field decays into 
the Nielsen-Olesen unstable modes.

\section{Effective Model of Monopoles and Their Production}
\label{3}

It is generally believed that
magnetic monopoles in QCD play the role in confining 
quarks. They condense in vacuum to form dual superconductors
in which color electric field is squeezed\cite{t,t1,dual} into a flux tube. Thus, quark confinement is realized.
We remind you that magnetic flux is squeezed in ordinary
superconductors where electrically charged Cooper pairs
condense. On the other hand, magnetically charged monopoles condense
in the dual superconductors so that electric flux is squeezed. 
In lattice gauge theories, 
we can see such a role of the magnetic monopoles with the use of maximally Abelian gauge\cite{max,max1,max2}.
Furthermore,
effective models of dual superconductors have been
explored\cite{koma} by using the lattice gauge theories. In the models
the magnetic monopoles are described by a complex scalar field with which 
dual gauge fields $A_{i,\rm eff}^d$
minimally couple.
Electric and magnetic fields are written in terms of the dual gauge fields 
such that
$E_i=\epsilon_{i,j,k}\partial_j A_{k,\rm eff}^d$
and $B_i=-\partial_0 A_{i,\rm eff}^d-\partial_i A_{0,\rm eff}^d$, respectively. 
( These electric and magnetic fields $E_i$ and $B_i$ are maximal Abelian components of 
SU(2) gauge fields, e.g. the third components $E_i^a=\delta^{a,3}E_i$ in color space.
Thus we can take $E_i$ and $B_i$ identical to 
the fields discussed in the previous section. Namely they are the fields
of the glasma. )
The electric field between q and $\bar{\rm{q}}$ is shown to form electric flux tube in the models. 
The parameters in the models\cite{dual} have been determined by fitting the profile of the flux tube to
the one obtained in the lattice gauge theories.  
The models describe not only electromagnetic properties of the monopoles and 
but also the behavior of the monopoles in a vaccum.
Although it is not obvious that the model is applicable to the analysis of the glasma, 
the use of the model in the analysis would be interesting attempt.
In this section we introduce an effective model of the monopoles and
calculate the production rate of the monopoles under background magnetic fields
as well as their back reaction to the magnetic field.

An effective model of the monopole field $\phi$ under background effective homogeneous gauge fields
$\vec{B}=(0,0,B_{\rm{eff}})$ and $\vec{E}=(0,0,E_{\rm{eff}})$
is given by

\begin{eqnarray}
\label{MO}
L_{\rm monopole} &=&|D_{\mu}^d\phi|^2+m^2|\phi|^2-\frac{\lambda}{4}|\phi|^4 \nonumber \\
&\simeq &
|\partial_t\phi|^2-|(\partial_3+ig_mA^d_{\rm{eff},3})\phi|^2
+(m^2-g_mE_{\rm{eff}})|\phi|^2
\end{eqnarray}
with $D_{\mu}^d\equiv \partial_{\mu}+ig_{\rm{m}}A_{\rm eff,\mu}^d$,
where $g_{\rm{m}}$ denotes magnetic charge ( $=2\pi/g$ ) of the monopoles.
The gauge fields $A_{\rm eff,3}^d$ is assumed to be spatially homogeneous.
We also assumed that the monopoles occupy the lowest Landau level
and neglected the quartic
interaction term, for simplicity.
The parameter $m$ approximately takes a value of the order of $1$GeV \cite{koma}.

The monopoles are spontaneously produced in a perturbative vaccum with $\langle \phi \rangle =0$ to form
a quark confining vacuum with their condensation $\langle \phi \rangle \neq 0$. The spontaneous creation
of the monopoles arises 
because the monopole field $\phi $ exponentially grow $\phi\propto \exp(mt)$ 
in the vacuum with $\langle \phi \rangle =0$.
This is the same as the case that the Nielsen-Olesen unstable modes $\phi_{NO}$ exponentially grow. 



First of all, we notice a similarity between the Lagrangian of the monopoles in eq(\ref{MO}) 
and that of the Nielsen-Olesen unstable modes
in eq(\ref{NO}).
Both excitations occupy the lowest Landau level under the electric ( magnetic ) field.
The monopoles possess the imaginary ``mass" $\sqrt{-m^2+g_mE_{\rm{eff}}}$ when $m^2>g_mE_{\rm{eff}}$,
 while the unstable modes
possess the imaginary ``mass" $\sqrt{-gB_{\rm{eff}}}$.
Both of them can be produced by the Schwinger mechanism; monopoles ( Nielsen-Olesen unstable modes ) are produced 
under magnetic ( electric ) field. 

The production rate of
the Nielsen-Olesen unstable modes has recently been obtained\cite{ita}
in the Schwinger mechanism .
Using the result,
we can easily obtain the production rate of the monopoles.
The production rate $\rm{R(NO)}$ of the Nielsen-Olesen unstable modes under the electric field $E_{\rm{eff}}$
has been found such that

\begin{equation}\label{R}
\rm{R(NO)} =\exp(\pi g B_{\rm{eff}}/g E_{\rm{eff}}),
\end{equation}
where the factor $\pi gB_{\rm{eff}}$ in $\rm{R(NO)}$ comes 
from the imaginary mass $\sqrt{-gB_{\rm{eff}}}$ in eq(\ref{NO}), that is,
$\pi gB_{\rm{eff}}=-\pi (\sqrt{-gB_{\rm{eff}}})^2$.
We should remember that
the production rate\cite{tanji} of a massive charged scalar particle with mass $M$ 
in the absence of magnetic fields
is given by $\exp(-\pi M^2/gE_{\rm{eff}})$, 
where we assumed that the transverse motion is frozen; $\vec{p_T}=0$.
( The transverse motion of the states in the lowest Landau level is also frozen. )
When we put $M=\sqrt{-gB_{\rm{eff}}}$ in the formula, we obtain the production rate $\rm{R(MO)}$.
Since the imaginary mass of the monopoles is given by $\sqrt{g_mE_{\rm{eff}}-m^2}$,
we can obtain the production rate $\rm{R(MO)}$ of the monopoles,

\begin{equation} 
\rm{R(MO)}=\exp\big(\pi (m^2-g_{\rm{m}}E_{\rm{eff}})/g_{\rm{m}}B_{\rm{eff}}\big),
\end{equation}
replacing the imaginary mass $\sqrt{-gB_{\rm{eff}}}$ by $\sqrt{g_mE_{\rm{eff}}-m^2}$
and replacing $gE_{\rm{eff}}$ by $gB_{\rm{eff}}$ in the formula $\rm{R(NO)}$.


It apparently seems unnatural that when the electric field in $\rm{R(NO)}$ vanishes i.e. $gE_{\rm{eff}}=0$, 
the production rate becomes infinity $\rm{R(NO)}=\infty$.
The fact contradicts the naive idea that when the electric field is absent,
the production of the charged particles does not arise. 
But we note that the amplitudes of the Nielsen-Olesen unstable modes indefinitely
grow $\phi_{NO}\propto \exp(t\sqrt{gB_{\rm{eff}}})$ when the electric field is absent.
It implies that the particle production indefinitely goes on.
On the other hand, when the electric field is present, the longitudinal momentum $\sim eE_{\rm{eff}}t+p_3$ 
becomes large owing to the acceleration by the electric field. 
Thus, even if the modes is unstable initially at $t=0$, 
the modes become stable when 
the energies $\sqrt{(eE_{\rm{eff}}t+p_3)^2-gB_{\rm{eff}}}$ become real.
Hence, the amplitudes $\phi_{NO}$ do not grow indefinitely when the electric field is present.
It implies that the particle production stops on the way.
This is the reason why $\rm{R(NO)}$ ( $\rm{R(MO)}$ ) 
is finite when $E_{\rm{eff}}\neq 0$ ( $B_{\rm{eff}}\neq 0$ ), while it
becomes infinite when $E_{\rm{eff}}= 0$ ( $B_{\rm{eff}}= 0$ ).
( The growth of the amplitudes $\phi_{NO}$, when the electric field is absent, stops owing to the four point
interaction $|\phi_{NO}|^4$. But the above result in eq(\ref{R}) does not take into account
the interaction. Thus, the amplitude grows indefinitely. ) 

Hereafter, for definiteness, we use the values of the parameters $m=0.7$ GeV, $Q_s=2$ GeV, $\alpha_s(Q_s)=1/3$ 
and $\sqrt{gB_{\rm{eff}}}=0.17\,Q_s$
( the value $\sqrt{gB_{\rm{eff}}}$ has been estimated using the results in the numerical
simulation\cite{berges4} ) as well as $E_{\rm{eff}}=B_{\rm{eff}}$. Then,
$g_mB_{\rm{eff}}=(2\alpha_s)^{-1}gB_{\rm{eff}}=(0.42\rm{GeV})^2$ with $\alpha_s\equiv g^2/4\pi$.
We should note that these values of $B_{\rm{eff}}$ and $E_{\rm{eff}}$ are the initial values
just after the production of the glasma. When we take into account back reaction of the monopoles
or Nielsen-Olesen unstable modes, they decrease with time. 


Numerically, we find that the production rate  
$\rm{R(MO)}\equiv\exp\big(\pi (m^2/g_{\rm{m}}B_{\rm{eff}}-1)\big)\simeq\exp(1.8\pi)$
of the monopoles is about 10 times larger than the rate $\rm{R(NO)}\equiv \exp(\pi)$ of
Nielsen-Olesen unstable modes; $\rm{R(MO)}/\rm{R(NO)}=\exp(0.8\pi)\simeq 12$. 
The production rate is the number of particles produced per unit volume.
Thus
the number of the monopoles produced is $10$ times larger than the number of the Nielsen-Olesen
unstable modes. 
In the estimation,
the back reaction of the produced particles to the background gauge fields
is not taken into account. As we will show in the next section,
when we take into account the back reaction,
we will see that the lifetime of the magentic field is 10 times shorter than that of the electric field.


\section{Rapid Decay of Color Magnetic Field into Monopoles}
\label{4}
Now, we consider the back reaction of the monopoles by using classical statistical
field theory\cite{csft2} and show the resultant rapid decay of the magnetic field.
 We only consider the production of the monopoles
in the lowest Landau level whose wavefunctions are given by

\begin{equation}
\phi\equiv (x_1-ix_2)^n\exp(-\frac{g_mE_{\rm{eff}}|z|^2}{4}+ip_3x_3),
\end{equation}
with $z\equiv x_1+ix_2$ and integer $n\ge 0$
where we used a gauge potential $\vec{A}_{\rm eff}^d=(-E_{\rm{eff}}x_2/2,E_{\rm{eff}}x_1/2,0)$.
The states are localized around $z=0$.
But, by taking the appropriate linear combination of the wavefunctions we
can form almost homogeneous field configurations
in the transverse plane.
Then, their magnetic currents $J_m(\phi)\equiv ig_m(\phi^{\dagger}D^d_3 \phi-(D^d_3\phi)^{\dagger}\phi)$ 
are also almost homogeneous so that 
the field $B_{\rm{eff}}$ governed by a Maxwell equation $\partial_t B_{\rm{eff}}=-J_m$ is
homogeneous. ( Here we have also assumed the homogeneity of the gauge field $B_{\rm{eff}}$ or 
the magnetic current $J_m$ even in the longitudinal 
direction. ) 
We assume that such field configurations are given by,

\begin{equation}
\label{sum}
\phi=\sum_{l=1\sim N}\psi_l(\vec{x}), \quad \psi_l(\vec{x})
\equiv \int dp_3 \,c(p_3)\exp(-\frac{g_mE_{\rm{eff}}|z-z_l|^2}{4}+ip_3x_3),
\end{equation} 
where $c(p_3)$ is a dimensionless function of the longitudinal momentum $p_3$ 
and $z_l\equiv x_{1.l}+ix_{2,l}$.
Each component $\psi_l(z)$ is approximately localized within the area $|z-z_l|<l_E$ 
and satisfies the condition, 
$\psi_l^{\dagger}\psi_{l'}\simeq \delta_{l,l'}|\psi_l|^2$ 
because we impose the condition that $|z_l-z_{l'}|\geq l_E\equiv \frac{1}{\sqrt{g_mE_{\rm{eff}}}}$.
( The monopole production does not affect the electric field $E_{\rm eff}$ so that
the value of $E_{\rm eff}$ keeps the initial value $\sqrt{gE_{\rm eff}}=0.17Q_s=0.34$GeV. 
Thus, the length parameter $l_E$ is constant. )
Namely, a component $\psi_l$ localized at $z=z_l$ is separated from the nearest neighbors 
approximately by the distance larger than $l_E$.
Furthermore, we assume that the number $N$ of the components $\psi_l$ 
is given by $N=kL^2/\,l^2_E$ where the parameter $k$ represents
how dense the transverse plane with the area $L^2$ is occupied by the fields $\psi_l$.
( Their number density is given by $N/L^2=k/l^2_E$. )
In order for our approximation to hold, we should take $k$ such that $k$ is not too
small ( the number density
is not too small ) to avoid inhomogeneity
in the transverse plane  
and not too large ( the number density is not too large ) to avoid over dense configuration of the field $\psi_l$.
For definiteness, we assume $k=1/10$ so that each component $\psi_l$ is separated from the others
by the distance equal to or larger than $\simeq 3l_E$. 
Then, the field configuration $\phi$ is 
approximately homogeneous.
This kind of the field configuration was analyzed\cite{ninomiya} to discuss so called
``spaghetti vacuum". ( In principle, we can determine the parameter $k$ by minimizing the energy of $\phi$
with respect to the variables $z_l$. Then,
the field configuration $\psi_l$ might form a lattice with appropriate lattice
spacing. The value of $k$ determined by such a procedure may be of the order of one. 
But, the precise value $k$ is not necessary to obtain our main results as you can see below. )

Using the field configuration in eq(\ref{sum}), we write down the energies of the monopoles and
the magnetic field,

\begin{eqnarray}
H_{\rm monopole} &=&\int d^3x \Bigg(\frac{1}{2}(\partial_tA_{\rm eff}^d)^2+|\partial_t\phi|^2+|D_i^d\phi|^2
-m^2|\phi|^2\Bigg)  \nonumber \\
 &\simeq & \int d^3x 
\Bigg(\frac{1}{2}(\partial_tA_{\rm eff}^d)^2+N(|\partial_t\psi|^2+|(i\partial_3-g_mA_{\rm eff}^d)\psi|^2
+(g_mE_{\rm{eff}}-m^2)|\psi|^2)\Bigg) \nonumber \\
 &=&N\int dx_3 \Bigg(\frac{l_E^2}{2k}(\partial_tA_{\rm eff}^d)^2+|\partial_tC|^2+|(i\partial_3-g_mA_{\rm eff}^d)C|^2
+(g_mE_{\rm{eff}}-m^2)|C|^2\Bigg)   \nonumber \\
 &=&NL\frac{l_E^2}{2k}(\partial_tA_{\rm eff}^d)^2+N\int dp 
\Bigg(|\partial_t C_p|^2+|(p+g_mA_{\rm eff}^d)C_p|^2+(g_mE_{\rm{eff}}-m^2)|C_p|^2\Bigg)
\end{eqnarray}
with $\int d^2x=L^2=Nl_E^2/k$ and $\int dx_3=L$,  
where 
\begin{eqnarray}
\psi &\equiv &C(x_3,t)\exp(-\frac{g_mE_{\rm{eff}}|z|^2}{4})\sqrt{\frac{g_mE_{\rm{eff}}}{2\pi}} \nonumber \\
 C(x_3,t)&\equiv &\int dp \,\frac{C_p}{\sqrt{2\pi}}\,\exp(ipx_3).
\end{eqnarray}

The color magnetic field is given by $B_{\rm{eff}}=\partial_0A_{\rm eff}^d$ in terms of 
the homogeneous dual gauge potential $A_{\rm eff}^d\equiv A_{\rm eff,3}^d$.

Using the Hamiltonian, we can derive the equation of motions of the fields $C_p(t)$
and $g_mA_d$,

\begin{eqnarray}
\label{eqm}
\partial_t^2C_p &=&(m^2-g_mE_{\rm{eff}})C_p-(p+g_mA_{\rm eff}^d)^2C_p   \nonumber \\
L\partial_t^2(g_mA_{\rm eff}^d)l_{\rm{E}}^2 &=&-\frac{2g_m^2k}{\pi}\int_{-\infty}^{+\infty} 
dp(p+g_mA_{\rm eff}^d)|C_p|^2,
\end{eqnarray}
with $k=1/10$,
where the second equation represents a dual Maxwell equation 
$\partial_t B_{\rm eff}=\partial_0^2A_{\rm eff}^d=-J_m$
with the magnetic current $J_m\equiv\frac{2g_mk}{Ll_{\rm{E}}^2\pi}\int_{-\infty}^{+\infty} dp(p+g_mA_{\rm eff}^d)|C_p|^2$.
It describes how the magnetic field decreases with the increase of the magnetic current $J_m$,
which is induced by the monopole production.
We note that the electric field $g_mE_{\rm{eff}}$ does not vary with time,
while the background magnetic field $g_mB_{\rm{eff}}(t)$ varies with time
owing to the back reaction of the
monopoles. 

To solve the equation, we need to impose initial conditions of
$A_{\rm eff}^d$ and $C_p$. The initial condition of $A_{\rm eff}^d$ is given by
$A_{\rm eff}^d(t=0)=0$ and $B_{\rm{eff}}(t=0)=\partial_tA_{\rm eff}^d(t=0)\equiv B_0$ 
where $B_0$ is the initial value of
the magnetic field; $g_mB_0=g_mE_{\rm{eff}}=(0.42\mbox{GeV})^2 \ll Q_s^2=(2\mbox{GeV})^2$.
On the other hand,
we should take initial conditions of the monopole field 
as shown by Dusling et al.\cite{csft}. By using the initial conditions, we can take into account
one loop quantum effects of the monopoles in our classical calculation.
The use of the initial conditions is the essence of the classical statistical field theory.

The initial conditions are given in the following,

\begin{eqnarray}
\label{ini}
C_p(t=0) &=&\frac{\exp(-\pi(1-\bar{m}^2)/8)}{(2g_mE_{\rm{eff}})^{1/4}}\Big(D_{(-1+i(\bar{m}^2-1))/2}
\Big(\frac{\sqrt{2}pe^{i\pi/4}}{\sqrt{g_mE_{\rm{eff}}}}\Big)d_p \nonumber \\
 &+&\bar{D}_{(-1+i(\bar{m}^2-1))/2}\Big(\frac{-\sqrt{2}pe^{i\pi/4}}{\sqrt{g_mE_{\rm{eff}}}}\Big)f_p\Big) \nonumber \\
\partial_tC_p(t=0) &=&\frac{\exp(-\pi(1-\bar{m}^2)/8)}{(2g_mE_{\rm{eff}})^{1/4}}\partial_t\Big(
D_{(-1+i(\bar{m}^2-1))/2}
\Big(\frac{\sqrt{2}(t\,g_mE_{\rm{eff}}+p)e^{i\pi/4}}{\sqrt{g_mE_{\rm{eff}}}}\Big)d_p  \nonumber \\
 &+&\bar{D}_{(-1+i(\bar{m}^2-1))/2}
\Big(\frac{\sqrt{2}(t\,g_mE_{\rm{eff}}-p)e^{i\pi/4}}{\sqrt{g_mE_{\rm{eff}}}}\Big)f_p\Big) 
\quad \mbox{as} \quad t\to 0
\end{eqnarray}
with parabolic cylinder function $D_{\nu}(z)$ and $\bar{m}^2\equiv m^2/g_mE\simeq 1.68$, 
where $d_p$ and $f_p$ denote Gaussian random variables satisfying

\begin{eqnarray}
\label{random}
\langle d_p\bar{d_q}\rangle &=&\langle f_p\bar{f_q}\rangle=\delta(p-q), \nonumber \\
\langle d_pd_q\rangle &=&\langle f_pf_q\rangle=\langle d_p\bar{f_q}\rangle=\langle f_p\bar{d_q}\rangle=
\langle \bar{d_p}\bar{d_q}\rangle=\langle \bar{f_p}\bar{f_q}\rangle=0. 
\end{eqnarray}
The initial conditions in eq(\ref{ini}) are derived 
from solutions $C_p(t \ll 1)$ which satisfy the equations (\ref{eqm})
in the limit $t\to 0$ where $A_{\rm eff}^d(t)\simeq B_0t=E_{\rm{eff}}t$.

The average $\langle\,Q(C_p(t),A_{\rm eff}^d(t))\,\rangle$ of a physical quantity $Q(C_p,A_{\rm eff}^d)$
over the Gaussian random variables $d_p$ and $f_p$ 
is taken 
after obtaining solutions of the equations (\ref{eqm}). We should note that the average $\langle |C_p|^2\rangle $
is proportional to $2\pi\delta(p=0)=\int dx_3 \exp(ix_3p)_{p=0}=L$. 
Hence, the factor $L$ in the left hand side of eq(\ref{eqm})
is cancelled with that of $\langle |C_p|^2\rangle $ in the right hand side. 
Thus, the length scale $L$ of the system in eq(\ref{eqm})  
does not cause any troubles.

By solving these equations with the initial condition $E_{\rm{eff}}=B_{\rm{eff}}(t=0)$, 
we can find how fast the magnetic field 
$B_{\rm{eff}}=\partial_tA_{\rm eff}^d$ decreases.
The decrease is caused by the production of the magnetic monopoles.

For comparison, we write down the corresponding equations for the Nielsen-Olesen unstable modes,
which describe the decrease of electric field $E_{\rm{eff}}=\partial_0A_{\rm eff} $ 
caused by the production of the unstable modes. ( The production of the unstable modes
does not affect the magnetic field $B_{\rm eff}$ so that the value of $B_{\rm eff}$ keeps the initial
value $\sqrt{gB_{\rm{eff}}}=0.17\,Q_s=0.34$GeV. )
The equations are in the following,

\begin{eqnarray}
\label{N}
\partial_t^2C_p^N &=&gB_{\rm{eff}}C_p^N-(p+gA_{\rm eff})^2C_p^N \nonumber \\ 
L\partial_t^2(gA_{\rm eff})l_{\rm{B}}^2 &=&-\frac{2g^2k}{\pi}\int_{-\infty}^{+\infty} dp(p+gA_{\rm eff})|C_p^N|^2,
\end{eqnarray}
with $l_{\rm{B}}\equiv \sqrt{1/gB_{\rm{eff}}}$,
where the correspondence between the monopole field $\phi$ and Nielsen-Olesen unstable modes $\phi_{NO}$
is obviously given by,

\begin{equation}
g_m \to g, \quad \psi \to \psi^N, \quad C_p\to C_p^N, \quad A^d_{\rm eff} \to A_{\rm eff},
\quad l_{\rm E} \to l_{\rm B} \quad
\mbox{and}  \quad E_{\rm eff}\,\, (B_{\rm eff}) \to B_{\rm eff} \,\,(E_{\rm eff}).
\end{equation} 
With these replacement, we can rewrite down the similar equations for the Nielsen-Olesen unstable modes 
to the equations for the monopoles.
Only the difference is that the equation for $C_p^N$ is given by
$\partial_t^2C_p^N =gB_{\rm{eff}}C_p^N-(p+gA_{\rm eff})^2C_p^N$, while the equation for $C_p$
is given by $\partial_t^2C_p =(m^2-g_mE_{\rm{eff}})C_p-(p+g_mA_{\rm eff}^d)^2C_p$.

Then, the initial conditions are in the following,

\begin{eqnarray}
C_p^N(t=0) &=&\frac{\exp(\pi/8)}{(2gB_{\rm{eff}})^{1/4}}\Big(D_{(-1+i)/2}
\Big(\frac{\sqrt{2}pe^{i\pi/4}}{\sqrt{gB_{\rm{eff}}}}\Big)d_p+
\bar{D}_{(-1+i)/2}\Big(\frac{-\sqrt{2}pe^{i\pi/4}}{\sqrt{gB_{\rm{eff}}}}\Big)f_p\Big) \nonumber \\
\partial_tC_p^N(t=0) &=&\frac{\exp(\pi/8)}{(2gB_{\rm{eff}})^{1/4}}\partial_t\Big(
D_{(-1+i)/2}
\Big(\frac{\sqrt{2}(t\,gB_{\rm{eff}}+p)e^{i\pi/4}}{\sqrt{gB_{\rm{eff}}}}\Big)d_p \nonumber \\
 &+&\bar{D}_{(-1+i)/2}
\Big(\frac{\sqrt{2}(t\,gB_{\rm{eff}}-p)e^{i\pi/4}}{\sqrt{gB_{\rm{eff}}}}\Big)f_p\Big) \quad \mbox{as} \quad t\to 0
\end{eqnarray}
with $A_{\rm eff}(t=0)=0$ and $\partial_tA_{\rm eff}(t=0)=E_{\rm{eff}}$,
where $d_p$ and $f_p$ are the random variables satisfying the above equations (\ref{random}).
The electric current $J_N$ is given such that
$J_N\equiv\frac{2gk}{Ll_{\rm{B}}^2\pi}\int_{-\infty}^{+\infty} dp(p+gA_{\rm eff})|C_p^N|^2$.
According to the Maxwell equation $\partial_tE_{\rm eff}=-J_N$ ( the second equation in eq(\ref{N}) ), the 
electric field $E_{\rm eff}=\partial_tA_{\rm eff}(t)$ varies with time.
( In the previous paper\cite{iwazaki} we have used different initial conditions for the Nielsen-Olesen unstable modes
from the ones used in the present paper. Using the initial conditions in the present paper,
we can correctly take into account quantum effects of the unstable modes. )

\vspace{0.2cm}
Before solving the above equations, we note that the average $\langle \,|C_p(t=0)|^2\rangle $ of
the quantity $|C_p(t)|^2 $ at $t=0$
is given such that

\begin{eqnarray}
\label{ini2}
\langle \,|C_p(t=0)|^2\,\rangle &=& \frac{L\exp(-\pi(1-\bar{m}^2)/4)}{2\pi(2g_mE_{\rm{eff}})^{1/2}}
\Big(|D_{(-1+i(\bar{m}^2-1))/2}
\Big(\frac{\sqrt{2}pe^{i\pi/4}}{\sqrt{g_mE_{\rm{eff}}}}\Big)|^2 \nonumber \\
&+&|\bar{D}_{(-1+i(\bar{m}^2-1))/2}\Big(\frac{-\sqrt{2}pe^{i\pi/4}}{\sqrt{g_mE_{\rm{eff}}}}\Big)|^2\Big).
\end{eqnarray}
Using this initial value
we will make an approximation in solving the equations. That is,
instead of taking the initial conditions in eq(\ref{ini}) 
involving the random variables $d_p$ and $f_p$, 
we take the following initial conditions which does not involve $d_p$ and $f_p$, 

\begin{eqnarray}
\label{ini3}
\tilde{C}_p(t=0)&=&\frac{\exp(-\pi(1-\bar{m}^2)/8)}{(2g_mE_{\rm{eff}})^{1/4}}\Big(|D_{(-1+i(\bar{m}^2-1))/2}
\Big(\frac{\sqrt{2}pe^{i\pi/4}}{\sqrt{g_mE_{\rm{eff}}}}\Big)|^2  \nonumber \\
&+&|\bar{D}_{(-1+i(\bar{m}^2-1))/2}\Big(\frac{-\sqrt{2}pe^{i\pi/4}}
{\sqrt{g_mE_{\rm{eff}}}}\Big)|^2\Big)^{1/2}\nonumber \\
\partial_t \tilde{C}_p(t=0) &=&\frac{\exp(-\pi(1-\bar{m}^2)/8)}{(2g_mE_{\rm{eff}})^{1/4}}\partial_t
\Big(|D_{(-1+i(\bar{m}^2-1))/2}\Big(\frac{\sqrt{2}(tg_mE_{\rm{eff}}+p)e^{i\pi/4}}
{\sqrt{g_mE_{\rm{eff}}}}\Big)|^2 \nonumber \\
&+&|\bar{D}_{(-1+i(\bar{m}^2-1))/2}\Big(\frac{\sqrt{2}(tg_mE_{\rm{eff}}-p)e^{i\pi/4}}{\sqrt{g_mE_{\rm{eff}}}}\Big)|^2\Big)^{1/2}
\quad \mbox{as} \quad t\to 0,
\end{eqnarray}
where we have rewritten $C_p$ such that $C_p=\sqrt{L/2\pi}\,\tilde{C}_p$.
In other words, the initial values $C_p(t=0)$ are given such that 
$C_p(t=0)=\sqrt{\langle |C_p(t=0)|^2\rangle }$; the right hand side of the equation is evaluated in eq(\ref{ini2}).
When we adopt our initial conditions, we do not need to take average over the random variables.
Obviously, the solutions obtained by using our simplified initial conditions in eq(\ref{ini3})
coincide with those obtained by using the proper initial conditions in eq(\ref{ini}) at least at 
the time $t=0$. 

According to the classical statistical field theory,
we have to take average over the Gaussian variables $d_p$ and $f_p$ involved in the solutions
via the initial conditions in eq(\ref{ini}). But tentatively we use our initial conditions in eq(\ref{ini3}) in order to 
greatly simplify our procedure of obtaining
solutions in eq(\ref{eqm}).
Although our procedure is a fairly rough approximation of the proper one, we will find that our results
are consistent with the ones given by the Schwinger mechanism.

\begin{figure}[htb]
  \centering
  \includegraphics*[width=2.5in]{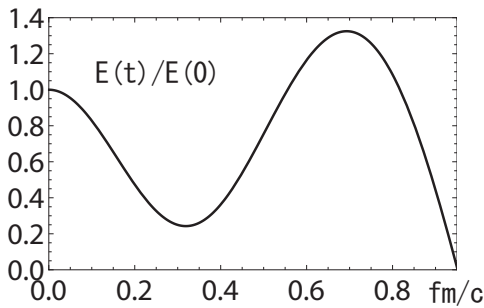}
  \caption{electric field decay with
the production of the Nielsen-Olesen unstable modes.}
\label{f1}
\end{figure}

\begin{figure}[htb]
  \centering
  \includegraphics*[width=2.5in]{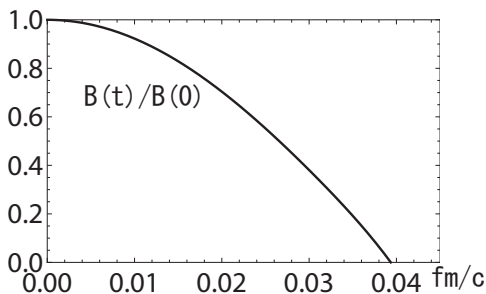}
  \caption{magnetic field decay with the production
of the magnetic monopoles.}
\label{f2}
\end{figure}

\begin{figure}[htb]
  \centering
  \includegraphics*[width=2.5in]{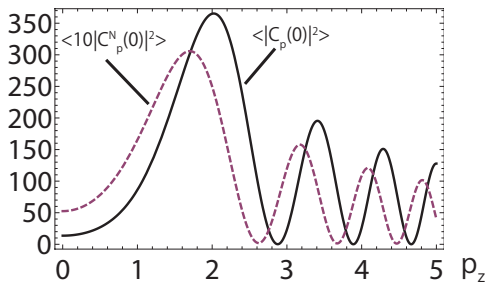}
  \caption{The initial amplitudes $\langle |C_p(t=0)|^2\rangle$ of the monopoles 
are 10 times larger than the initial amplitudes $\langle |C_p^N(t=0)|^2\rangle$ 
of Nielsen-Olesen unstable modes. The horizontal axis denotes momentum nomalized such as 
$p_z\equiv p/\sqrt{gB_{\rm{eff}}}$  for $\langle |C_p^N(t=0)|^2\rangle $ and 
$p_z\equiv p/\sqrt{g_mE_{\rm{eff}}}$ for $\langle |C_p(t=0)|^2\rangle$.}
\label{f3}
\end{figure}

In Fig.\ref{f1} we show that the background color electric field decreases with the production of
the Nielsen-Olesen unstable modes. Similarly, in Fig.\ref{f2} we show 
that the background color magnetic field decreases with the production of
the magnetic monopoles.
We can see that the magnetic field decreases more rapidly than the electric field.
The ratio of the lifetime $\rm T_E$ of the electric field ( $E_{\rm eff}(\rm T_E)=0$ ) to the lifetime
$\rm T_M$ of the magnetic field ( $B_{\rm eff}(\rm T_B)=0$ ) is given by $\rm T_E/T_M\simeq 24$.   
We can show that the difference in the lifetimes comes from the difference 
in the initial conditions. Namely, the
initial amplitude $C_p(t=0)$ of the monopole field is much larger than
that of the Nielsen-Olesen unstable modes $C_p^N(t=0)$
( Fig.\ref{f3} ). 
Physically, it means that
the magnetic current is much larger than the electric current in the early stage $t\sim 0$;
$J\propto |C_p|^2$. The larger magnetic current gives rise to the more rapid decrease of the 
magnetic field. 

Although the field amplitude $C_{p=0}(t)$ may begin to grow exponentially such as $C_{p=0}\propto \exp(tm)$
for $t>m^{-1}\simeq 0.3 \rm{fm/c}$,
the magnetic field vanishes before the start of the exponential growth. Thus, the exponential
growth of the amplitude does not contribute to the decrease of the magnetic field.
It decreases mainly due to the large amplitude of the monopole field
in the very early stage. 

The result concerning to the lifetimes is consistent with the one obtained above;
the production rate $\rm{R(MO)}$ of the monopoles in the Schwinger mechanism
is 10 times larger than that $\rm{R(NO)}$ of Nielsen-Olesen
unstable modes ( $\rm{R(MO)}/\rm{R(NO)}\simeq 12$. )

Here we make some comments on the magnetic ( $J_m$) and electric ( $J_N$ ) currents, for example, 
$J_m=\frac{2g_mk}{Ll_{\rm{E}}^2\pi}\int_{-\infty}^{+\infty} dp(p+g_mA_{\rm eff}^d)|C_p|^2$ in eq(\ref{eqm}).
The first comment is 
that $J_m$ vanishes at $t=0$ because $g_mA_{\rm eff}^d(t=0)=0$ and the integrand is antisymmetric in the variable $p$;
$-p|C_{-p}(t=0)|^2=-p|C_p(t=0)|^2$.
Similarly, the electric current $J_N$ vanishes at $t=0$.
Then, the current begins to flow after the spontaneous production of the magnetic monopoles.
The amount of the current is determined 
by the large amplitude of the monopole field $|C_p(t=0)|$ ( $ \gg |C_p^{NO}(t=0)|$ ).
This initial large amount of the monopole current causes the magnetic field decrease rapidly.
The second one is
concerned with the dependence of the currents on the parameter $k$. 
Because $J_N\propto k$ and $J_m\propto k$, the parameter $k$ can be absorbed by 
the amplitudes $C_p$ and $C_p^N$ in eq(\ref{eqm}) and eq(\ref{N}) with the replacement 
such as $C_p\to k^{-1/2}C_p$ and $C_p^N\to k^{-1/2}C_p^N$. Then, the equations of motion
do not involve the parameter $k$.
But, both of the initial amplitudes $C_p(t=0)$ and $C_p^N(t=0)$
acquire the factor $k^{1/2}$. Thus, the ratio of $|C_p(t=0)|^2$ to $|C_p^N(t=0)|^2$
is independent of $k$. This large value of the ratio $|C_p(t=0)|^2/|C_p^N(t=0)|^2\sim 10$
gives rise to the dominance of the monopole production in the glamsa decay.  
Therefore, although there is the ambiguity of the value $k$ used in our calculations, 
our result of the rapid glasma decay into the monopoles still holds.  
( Here we have assumed that the rapid decay of the magnetic field into the monopoles
also leads to the rapid decay of the electric field. This is because the flux tube of the electric
field expands\cite{instability1} into the direction perpendicular to the tube, 
which generates magnetic field according to
the dual Faraday's law. Then the magnetic field may also decays into the monopoles. Thus, it turns out that 
the electric field also decays producing the magnetic monopoles.   )

Our results have been obtained using the simplified initial conditions.
The conditions are not proper ones. But using the initial conditions
we obtain the same initial large amplitude as the one $\langle |C_p(t=0)|^2\rangle $
obtained by using the proper initial conditions.
That is, Both initial conditions give rise to the large initial amplitudes of
the monopole field, compared with the initial amplitudes of the Nielsen-Olesen unstable modes.
As we have shown, these initial large amplitudes cause the rapid decay 
of the magnetic field. Hence, 
our conclusion of the rapid decay of the glasma into the monopoles
may be reliable even if the simplified initial conditions are adopted.

\section{Discussions and Conclusions}
\label{5}

We have shown that the glasma rapidly decays into magnetic monopoles.
Then, it is natural to ask how the monopole gas leads to the thermalized QGP.
Classical solutions of color magnetic monopoles
are unstable unless their stability is guaranteed topologically.
Indeed, there are no solutions of stable magnetic monopoles in real QCD.
It has been shown\cite{coleman} that the growth rates of unstable modes of gluons around 
classical monopoles
can be infinitely large; the rates
 are proportional to the logarithm of the volume in 
the system. The fact indicates that the monopoles rapidly decay into gluons
even if they are produced. Furthermore, they couple strongly with gluons
because the smaller the $g$, the larger the coupling $g_m=2\pi/g$. Therefore,
thermalized QGP would be generated immediately
after the decay of the glasma into the monopoles.
 


In our analyses we have used a classical statistical field theory technique
in order to take into account the one loop quantum effects of the monopoles and the unstable modes.
To comfirm the validity of the technique,
we need to clarify whether or not the higher order effects of these excitations change our results.

The model of the monopoles used in the present paper is 
a model of dual superconductor in which quark confinement is realized.
The model is an effective model of the monopoles.
It is not clear that the model is applicable to the analysis of the glasma decay.
But, our results suggest that the monopoles play significant roles in
the glasma decay. 
Therefore, it is desirable to perform
more rigorous treatment of the monopoles in the analysis of the glasma.

\vskip2pc

\vspace{0.2cm}
The author
expresses thanks to Prof. T. Kunihiro and Dr. N. Tanji
for their useful comments.




\begin{thebibliography}{99}
\bibitem{cgc}E. Iancu, A. Leonidov and L. McLerran, hep-ph/0202270.
\bibitem{cgc1}E. Iancu and R. Venugopalan, hep-ph/0303204.
\bibitem{hirano}T. Hirano and Y. Nara, Nucl. Phys. {\bf A743} (2004) 305; J. Phys. {\bf G30} (2004) S1139.
\bibitem{ven}P. Romatschke and R. Venugopalan, Phys. Rev. Lett. {\bf 96} (2006) 062302;
Phys. Rev. {\bf D74} (2006) 045011.
\bibitem{fuku}K. Fukushima and F. Gelis, Nucl. Phys. {\bf A874} (2012) 108.
\bibitem{berges1}J. Berges and S. Schlichting, Phys. Rev. {\bf D87} (2013) 014026.
\bibitem{berges}J. Berges, S. Scheffler and D. Sexty, Phys. Rev. {\bf D77} (2008) 034504.
\bibitem{kunihiro1}T. Kunihiro, B. Muller, A. Ohnishi, A. Schafer, T. T. Takahashi and A Yamamoto,
Phys.Rev. {\bf D82} (2010) 114015. 
\bibitem{kunihiro2}H. Iida, T. Kunihiro, B. Muller, A. Ohnishi, A. Schafer and T. T. Takahashi, hep-ph/13041807. 
\bibitem{nielsen}N.K. Nielsen and P. Olesen, Nucl. Phys. {\bf B144} (1978) 376.
\bibitem{instability}A. Iwazaki, Phys. Rev. {\bf C77} (2008) 034907; Prog. Theor. Phys. {\bf 121} (2009) 809.
\bibitem{instability1}H. Fujii and K. Itakura, Nucl. Phys. {\bf A809} (2008) 88.
\bibitem{instability2}H. Fujii, K. Itakura and A. Iwazaki, Nucl. Phys. {\bf A828} (2009) 178.
\bibitem{berges2}J. Berges, S. Scheffler, S. Schlichting and D. Sexty, Phys. Rev. {\bf D85} (2012) 034507. 
\bibitem{coleman}S. Coleman, "The magnetic monopole 50 years later" 
in The Unity of the Fundamental Interactions (1983), A. Zichichi, editor. 
\bibitem{t}G.t'Hooft, High Energy Physics, edited by A. Zichichi ( Editorice Compositori, Bologna,1975 ).
\bibitem{t1}S. Mandelstam, Phys. Rep {\bf 23} (1976) 245.
\bibitem{ie}Z.F. Ezawa and A. Iwazaki, Phys. Rev. {\bf D25} (1982) 2681.
\bibitem{s}T. Suzuki and I. Yotsuyanagi, Phys. Rev. {\bf D42} (1990) 4257.
\bibitem{dual}see a review article, G. Ripka, lecture notes in physics {\bf 639}, Springer.
\bibitem{max}T. Suzuki and I. Yotsuyanagi, Phys. Rev. {\bf D42} (1990) 4257.
\bibitem{max1}J.D. Stack, S.D. Neiman and R.J. Wensley, Phys. Rev. {\bf D50} (1994) 3399.
\bibitem{max2}K. Amemiya and H. Suganuma, Phys. Rev. {\bf D60} (1999) 114509.
\bibitem{koma}Y. Koma, M. Koma, E.-M. Ilgenfritz and T. Suzuki, 
Phys. Rev. {\bf D68}, (2003) 114504.
\bibitem{kondo}K.I. Kondo, Phys. Lett. {\bf B600} (2004) 287.
\bibitem{sch}J. Schwinger, Phys. Rev. {\bf 82} (1951) 664.
\bibitem{ita}N. Tanji and K. Itakura, Phys. Lett. {\bf B173} (2012) 112.
\bibitem{csft}K. Dusling, F. Gelis and R. Venugopalan, Nucl. Phys. {\bf A872} (2011) 161.
\bibitem{csft1}K. Dusling, T. Epelbaum, F. Gelis and R. Venugopalan, Phys. Rev. {\bf D86} (2012) 085040.
\bibitem{csft2}F. Gelis and N. Tanji, Phys. Rev. {\bf D87} (2013) 125035.
\bibitem{vis}J. Liao and E. Shuryak, Phys. Rev. Lett. {\bf 101} (2008) 162302. 
\bibitem{tanji}N. Tanji, Annals. Phys. {\bf 324} (2009) 1691 ( see the references therein ).
\bibitem{berges4}J. Berges, D. Gelfand, S. Scheffler and D. Sexty, Phys. Lett. {\bf B677} (2009) 210.
\bibitem{ninomiya}H.B. Nielsen and M. Ninomiya, Nucl. Phys. {\bf B156} (1979) 1.
\bibitem{iwazaki}A. Iwazaki, Phys. Rev. {\bf C87} (2013) 024903.
\end{thebibliography}
\end{document}